\documentclass[sigconf]{acmart}

\usepackage{enumitem}  
\usepackage{multirow}
\usepackage{cleveref}

\AtBeginDocument{%
  }

\setcopyright{acmlicensed}
\copyrightyear{2025}
\acmYear{2025}
\setcopyright{acmlicensed}\acmConference[KDD '25]{Proceedings of the 31st ACM SIGKDD Conference on Knowledge Discovery and Data Mining V.1}{August 3--7, 2025}{Toronto, ON, Canada}
\acmBooktitle{Proceedings of the 31st ACM SIGKDD Conference on Knowledge Discovery and Data Mining V.1 (KDD '25), August 3--7, 2025, Toronto, ON, Canada}
\acmDOI{10.1145/3690624.3709438}
\acmISBN{979-8-4007-1245-6/25/08}

\begin{document}

\title{Multi-granularity Interest Retrieval
and Refinement Network for Long-Term User Behavior Modeling in CTR Prediction}

\author{Xiang Xu}
\email{demon@mail.ustc.edu.cn}
\affiliation{%
  \institution{University of Science and Technology of China, State Key Laboratory of Cognitive Intelligence}
  \city{Hefei}\country{China}
}

\author{Hao Wang}
\authornote{Corresponding author}
\email{wanghao3@ustc.edu.cn}
\affiliation{%
  \institution{University of Science and Technology of China, State Key Laboratory of Cognitive Intelligence}
  \city{Hefei}\country{China}
}

\author{Wei Guo}
\email{guowei67@huawei.com}
\affiliation{%
  \institution{Huawei Noah's Ark Lab} 
  \country{Singapore}
}

\author{Luankang Zhang}
\email{zhanglk5@mail.ustc.edu.cn}
\affiliation{%
  \institution{University of Science and Technology of China, State Key Laboratory of Cognitive Intelligence} 
  \city{Hefei}\country{China}
}

\author{Wanshan Yang}
\email{wanshan.yang@colorado.edu}
\affiliation{%
  \institution{Consumer Cloud Service Interactive Media BU, Huawei} 
  \city{Shenzhen}\country{China}
}


\author{Runlong Yu}
\email{yrunl@mail.ustc.edu.cn}
\affiliation{%
  \institution{University of Science and Technology of China, State Key Laboratory of Cognitive Intelligence}
  \city{Hefei}\country{China}
}

\author{Yong Liu}
\email{liu.yong6@huawei.com}
\affiliation{%
  \institution{Huawei Noah's Ark Lab} 
  \country{Singapore}
}

\author{Defu Lian}
\email{liandefu@ustc.edu.cn}
\affiliation{%
  \institution{University of Science and Technology of China, State Key Laboratory of Cognitive Intelligence}
  \city{Hefei}\country{China}
}

\author{Enhong Chen}
\email{cheneh@ustc.edu.cn}
\affiliation{%
  \institution{University of Science and Technology of China, State Key Laboratory of Cognitive Intelligence} 
  \city{Hefei}\country{China}
}

\renewcommand{\shortauthors}{Xiang Xu, et al.}

\begin{abstract}

Click-through Rate (CTR) prediction is crucial for online personalization platforms. Recent advancements have shown that modeling rich user behaviors can significantly improve the performance of CTR prediction. 
Current long-term user behavior modeling algorithms predominantly follow two cascading stages. The first stage retrieves subsequence related to the target item from the long-term behavior sequence, while the second stage models the relationship between the subsequence and the target item. 
Despite significant progress, these methods have two critical flaws. First, the retrieval query typically includes only target item information, limiting the ability to capture the user's diverse interests. 
Second, relational information, such as sequential and interactive information within the subsequence, is frequently overlooked. Therefore, it requires to be further mined to  more accurately model user interests.

To this end, we propose Multi-granularity Interest Retrieval and Refinement Network (MIRRN). 
Specifically, we first construct queries based on behaviors observed at different time scales to obtain subsequences, each capturing users’ interest at various granularities. 
We then introduce an noval multi-head Fourier
transformer to efficiently learn sequential and interactive information within the subsequences, leading to more accurate modeling of user interests. 
Finally, we employ multi-head target attention to adaptively assess the impact of these multi-granularity interests on the target item.
Extensive experiments have demonstrated that MIRRN significantly outperforms state-of-the-art baselines. Furthermore, an A/B test shows that MIRRN increases the average number of listening songs by 1.32\% and the average time of listening songs by 0.55\% on the Huawei Music App. The implementation code is publicly available at \textcolor{blue}{\url{https://github.com/USTC-StarTeam/MIRRN}}.

\end{abstract}


\maketitle

\section{Introduction}
Click-Through Rate (CTR) prediction is a crucial task in modern online personalization platforms, such as recommendation systems and online advertising~\cite{oa1, oa2}. Its objective is to estimate the probability of a user clicking on a specific item within a given contextual situation~\cite{ctr01, deepfm}. Recent advancements in CTR prediction have shown that user behavior modeling which incorporates the user's historical behavior sequence to learn implicit interests results in significant performance gain~\cite{ ctr3, DIN, DIEN}. As online platforms evolve, the length of user behavior sequences has experienced a surge.For instance, on Taobao, one of China's largest e-commerce platforms, over 23\% of users have recorded more than 1,000 clicks in the past five months alone~\cite{lifelong}.
Effectively modeling this massive and intricate user behavior data has emerged as a focal point for both academia and industry. 

Due to the strict online inference time constraints, previous methods~\cite{DIN, DIEN, SAS, bert} considering their high time complexity, can only utilize the most recent behaviors with the sequence length seldom exceeding 100.
This limitation leads to suboptimal results.
To address this, several models have been proposed to leverage long behavior sequences for modeling user interests. Most of these models follow two cascading stages~\cite{SIM}. 
In the first stage, the General Search Unit (GSU) identifies a small number of behaviors most relevant to the target item from thousands of long-term behaviors using fast and coarse search methods. In the second stage, the Exact Search Unit (ESU) captures the relationships between behaviors in these subsequences and the target item to precisely learn user interests. The most critical difference among these models such as SIM~\cite{SIM}, UBR4CTR~\cite{UBR4CTR}, ETA~\cite{ETA}, SDIM~\cite{SDIM}, UBCS~\cite{UBCS} and TWIN~\cite{TWIN}, lies in the different search strategies adopted by GSU.

\begin{figure}[t]
  \centering
  \includegraphics[width=\linewidth]{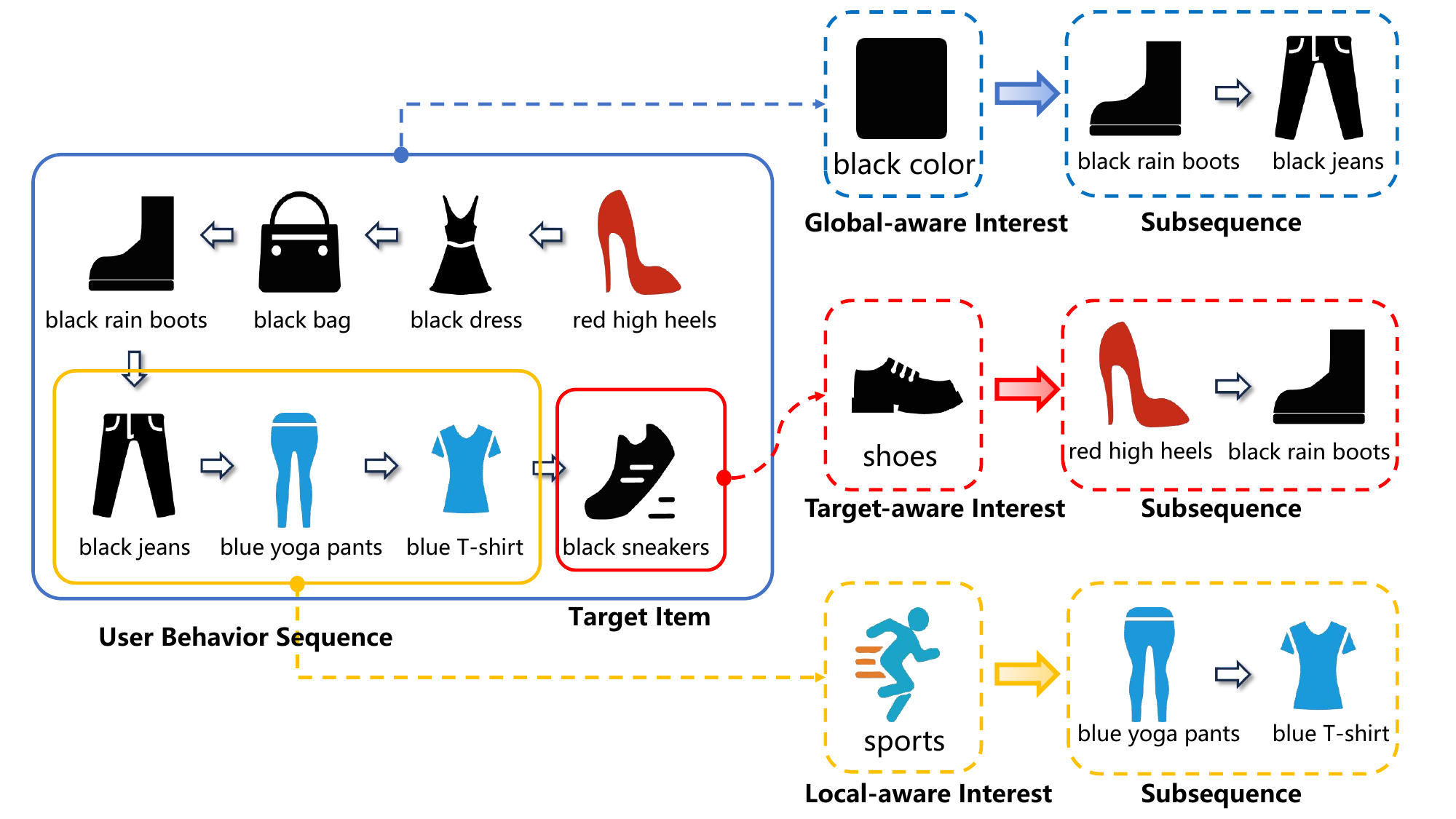}
  \caption{The left side of the picture depicts a user's behavior sequence, with black sneakers as the target item. The middle column represents the microcosms of the user's multi-granularity  interests, corresponding to the diverse queries we aim to construct. The right column displays the subsequences related to the multi-granularity interests.}
  \label{fig:intro}
  \Description{}
\end{figure}

Despite significant improvements in performance and efficiency, most two-stage models~\cite{UBR4CTR, SIM, ETA} focus on limited contextual information in the first stage and neglect to mine the internal relations of subsequences in the second stage.
In the first stage, we argue that the retrieval process, where the retrieval query contains only the target item information, accounts solely for target-aware interest.
However, if the retrieval query includes relevant information about behaviors at different time scales, more diverse user interests can be mined. Ignoring these "multi-granularity interests", weakens the representation power of the learned user interest.
For example, consider the user behavior sequence shown in Figure ~\ref{fig:intro}. Most models capture target-aware interests, favoring "shoes" with a high probability when focusing on the target item "black sneakers". However, we believe that coarser-grained user interests, such as "local-aware" and "global-aware" interests, also impact the final prediction results. Recent clicks on "blue yoga pants" and "blue T-shirt" indicate the user's local-aware interest in "sports". Additionally, broader user behavior sequence indicates the user's global interest in "black" items. In this example, these interests in shoes, sports, and black all positively influence the click rate on black sneakers. 
In the second stage, we believe that the relational information in the subsequence, such as sequential and interactive information, is as valuable as that in the original sequence~\cite{SAS}. It needs to be further mined to model more accurate user interests.

In fact, we identify two key challenges in the long-term behavior modeling that require further investigation:
(1) Effectively capturing diverse user interests under stringent online inference time constraints presents a non-trivial challenge.
(2) Efficiently mining the sequential and interactive information in the subsequences to learn more accurate user interests remains a critical challenge.

To address the aforementioned challenges, we propose \textbf{M}ulti-granularity \textbf{I}nterest  \textbf{R}etrieval and \textbf{R}efinement \textbf{N}etwork \textbf{(MIRRN)}. Firstly, we propose Multi-granularity Interest Retrieval Module (MIRM) to capture diverse user interests. MIRM constructs queries based on the behaviors observed at different time scales, which contain the target, local, and global information of the behavior sequence, respectively. Using these queries, MIRM adopts a fast search method based on SimHash to handle behavior sequences of thousands of lengths, obtaining subsequences that each contains interest of different granularities.
Secondly, we propose Behavior Sequence Refinement Module (BSRM) to extract relational information in subsequences. To incorporate sequential information and efficiently mine interactive information, BSRM employs Target-Aware Position Encoding (TAPE) and Multi-Head Fourier Transformer (MHFT), respectively.
MHFT utilizes the fast Fourier transform (FFT) to replace the complex convolution operation in the time domain, thereby enabling interaction with a low time complexity of $O(nlogn)$. Additionally, MHFT introduces matrix multiplication and multi-head mechanism to enhance interaction and focus on information in different subspaces, while greatly reducing the number of parameters.
Finally, we propose Multi-granularity Interest Activation Module (MIAM) using the multi-head target attention to adaptively learn the influence of different granular interests on the target item, obtaining the final user representation.

In summary, our main contributions are as follows:

\begin{itemize}[leftmargin=*,align=left]
\item In our proposed MIRRN, MIRM construct diverse queries based on behaviors observed at different time scales to obtain subsequences. This approach enables MIRM to effectively capture multi-granularity interests.
To the best of our knowledge, we are the first to successfully capture, refine, and integrate user’s multi-granularity interests in long-term user behavior modeling.
\item In our proposed MIRRN, BSRM adopts the target-ware position encoding and the noval multi-head Fourier transformer (MHFT) to mine sequential and interactive information.
MHFT innovatively utilizes the fast Fourier transform, matrix multiplication and multi-head mechanism to efficiently and comprehensively mine interactive information.
\item Extensive experiments have demonstrated that MIRRN achieves better performance compared to state-of-the-art baselines. We verify the validity of each module through comprehensive ablation studies. Furthermore, an online A/B testing conducted on the Huawei Music App further reinforces the effectiveness of  MIRRN in real industrial scenarios.
\end{itemize}

\section{Related Works}
\subsection{Click-Through-Rate Prediction}

Early CTR models primarily concentrate on exploiting feature interactions. Pioneering studies are relatively shallow such as factorization machines(FM)~\cite{FM} and its variants~\cite{FM1, FM2, FM3}. With the great success of deep learning, various studies such as Wide\&Deep~\cite{wide}, PNN~\cite{PNN}, DeepFM~\cite{deepfm} and DCN~\cite{DCN, DCN2}, successfully combine deep and shallow models. Moreover, several models~\cite{AFM, NFM, xdeepfm, CELS, celslake} have explored more complex neural networks, such as xDeepFM~\cite{xdeepfm} further introduces compressed interaction networks to explicitly model high-order feature interactions.

With the increasing demand for personalization, researchers are focusing on user behavior modeling~\cite{DIN, yin2023apgl4sr, shen2024exploring, DIEN, TIN}, which derives the latent representation of users' interest from their behavior sequences. YoutubeDNN~\cite{youtube} performs mean pooling on the entire user sequence. DIN~\cite{DIN} proposes an attention mechanism to identify target-relevant interests. DIEN~\cite{DIEN} introduces a novel GRU~\cite{gru} with an attention update gate (AUGRU) to capture the evolving process of user interests.
DSIN~\cite{DSIN} observes that user behaviors are highly homogeneous in each session, and heterogeneous cross sessions.
DMIN~\cite{DMIN} and MIND~\cite{mind} employ self-attention mechanisms and capsule networks, respectively, to represent user interests as multiple vectors. 
CAN~\cite{CAN} captures the feature co-action by directing user behavior features into a compact multi-layer perceptron generated by target features.

\subsection{Long-Term User Behavior Modeling}
As user behavior modeling~\cite{ han2024efficient, yin2024dataset, xie2024bridging, DIN} has demonstrated remarkable performance in industrial applications, researchers start to focus on modeling increasingly longer behaviors. 
Due to strict limitations on inference time and model parameter count, DIN-like models~\cite{DIN,DIEN,DSIN} are less suitable for long sequence modeling.

Recently, various models~\cite{MIMN, SIM, UBCS, ETA, SDIM, TWIN} have been proposed to leverage long behavior sequences for modeling user interests.
MIMN~\cite{MIMN} stores user behaviors in the memory matrix of the User Interest Center (UIC). 
However, MIMN does not pay much attention to the target item information.
SIM~\cite{SIM} introduces a two-stage cascading framework to overcome above issue. 
In the first stage, the General Search Unit (GSU) uses a rapid and coarse search method to retrieve several behaviors related to the target item from thousands of long-term behaviors, forming subsequences. In the second stage, the Exact Search Unit (ESU) captures the relationships between behaviors in these subsequences and the target item to precisely learn user interests.
The most critical difference between later models lies in the different search strategies adopted by GSU.
For example, SIM Hard~\cite{SIM} selects behaviors in the same category as the target item. 
SIM Soft~\cite{SIM} employs the inner product of pre-trained item embeddings as the relevance measure. 
UBR4CTR~\cite{UBR4CTR} utilizes BM25 algorithm to search for related behaviors. 
ETA~\cite{ETA} proposes an end-to-end model that leverages locality-sensitive hashing~\cite{simmm} (LSH) to encode item embeddings and retrieve relevant items via Hamming distance. 
SDIM~\cite{SDIM} directly gathers behavior items that share the same hash signature with the candidate item, then performs a linear aggregation on them. 
UBCS~\cite{UBCS} adopts a sampling method that considers relevance and temporal information and accelerates the sampling process through clustering.
TWIN~\cite{TWIN} designed the Consistency-Preserved GSU (CP-GSU) by behavior feature splitting, using the same target behavior correlation measure as ESU.
Additionally, several approaches attempt to directly shorten the length of user behaviors. 
ADFM~\cite{ADFM} uses hierarchical aggregation to compress raw behavior sequences.
DGIN~\cite{DGIN} groups behaviors using a relevant key to remove abundant items.
Based on TWIN, TWIN-V2~\cite{TWIN2} further employs a hierarchical clustering method to group items with similar characteristics into a cluster.

Nevertheless, these models overlook the diversity of user interests and the relationships among behaviors in the subsequences. Our MIRRN focuses on learning multi-granular user interests and efficiently mining sequential and interactive information .


\section{Methodology}
\begin{figure*}[h]
  \centering
  \includegraphics[width=\linewidth]{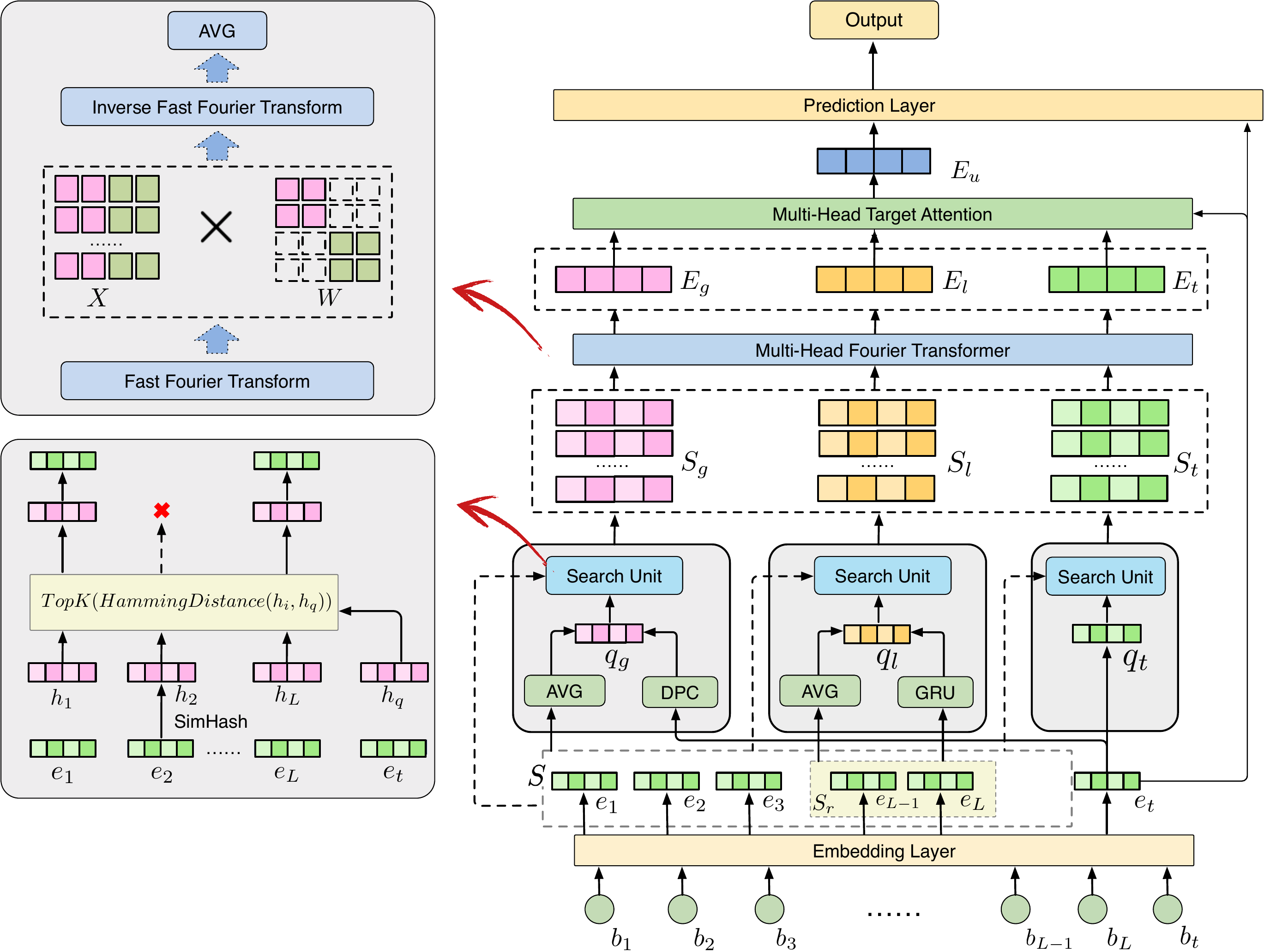}
  \caption{The Architecture of our proposed MIRRN.}
  \label{model}
  \Description{}
\end{figure*}

The framework of MIRRN is depicted in Figure~\ref{model}. Our model comprises Embedding Layer, Multi-granularity Interest Retrieval Module, Behavior Sequence Refinement Module, Multi-granularity Interest Activation Module and Prediction Layer. The details of these modules will be introduced in the subsequent subsections.

\subsection{Embedding Layer}
Depending on the nature of the input features, we utilize diverse embedding techniques. We categorize input features into two main types: categorical and numeric features. We perform one-hot encoding and multi-hot encoding on categorical features. For numerical features, we first assign them to different numerical buckets and then perform one-hot encoding to distinguish these buckets.
It's typical to map all one-hot feature vectors into a lower-dimensional latent vector space. Specifically, we define the embedding vector of the $i$-th behavior in the user behavior sequence $b_i$ as $e_i \in \mathbb{R}^{d}$, where $d$ represents the embedding size. We consolidate all the embedding vectors of user behavior items into a matrix $S \in \mathbb{R}^{L \times d}$, representing the embedding representation of the user behavior sequence:
\begin{equation}
    S = [e_1; e_2; \dots; e_L].
\end{equation}
where $L$ denotes the length of the user behavior sequence. We denote the embedding of the target item as $e_t$. For other features, we abbreviate them as $E_{other}$.


\subsection{Multi-granularity Interest Retrieval Module}       
Considering the strict limitations on inference time and the number of model parameters for long sequence modeling, we follow previous work~\cite{ETA} using a simple and fast search method to filter out relevant items from the user's long-term behavior sequence to obtain subsequence (SBS). We then accurately model the user's interests within SBS to reduce training and inference costs. However, in previous work, the query in the retrieval stage considers only the target information, resulting in SBS that contain only the user's target-relevant interest. We construct queries based on behaviors observed at different time scales, containing target, local, and global information from the behavior sequence, respectively. These queries serve as microcosms of the user's diverse interests, allowing the retrieved SBSs to contain interests of different granularities. Below, we will introduce the methods for constructing different queries and the fast search units in detail.
\subsubsection{Target-Aware Search Unit (TASU)}
The information contained in the target item is crucial for accurately predicting the user's click rate on it. Therefore, in this unit, we directly use the embedding of the target item as the query $q_{t}$.
Using this query, we can obtain the SBS $S_{t} \in \mathbb{R}^{K \times d}$ through a fast search method. $K$ is the predefined length of the SBS. $S_{t}$ contains the user's most fine-grained interest, which we refer to as target-aware interest. 
\begin{equation}
    q_{t} = e_t,
\end{equation}
\begin{equation}
    S_{t} = Search(S, q_{t}),
\end{equation}
where the $Search$ function receives the behavior sequence and query as input, and outputs related SBS. 

In our paper, we adopt a Simhash-based search method for its simplicity and efficiency.
The SimHash algorithm takes an item's embedding vector as input and transforms it through a hash function into a binary signature as output. 
Simhash adheres to the locality-sensitive property: if the input vectors are similar, then the output binary signatures will also be similar.
Specifically, we employ a fixed random matrix $H \in \mathbb{R}^{d \times m}$ where the elements are generated randomly around 0. Each row of $H$ is served as a random hash function.
Through matrix multiplication and the sign function, item embedding $e_i$ and query embedding $q_{t}$ can be transformed into m-dimensional binary signatures $h_i, h_q \in \mathbb{R}^{1 \times m}$.
\begin{equation}\label{E4}    
    h_i = SimHash(e_i) = ReLU(Sign(e_i H)),
\end{equation}
\begin{equation}\label{E5}    
    h_{q_t} = SimHash(q_{t}) = ReLU(Sign(q_{t} H)),
\end{equation}

Due to the locality-sensitive property, the similarity between an item and a query can be replaced by the similarity between their binary signatures. 
Therefore, we can use the Hamming distance between the query and item binary signatures as a similarity indicator instead of the inner product of the embedding vectors, thereby reducing the time complexity of calculating similarity.
By comparing the Hamming distances of all behaviors to the query, we obtain the indices $I_t \in \mathbb{N}^{K}$ of the $K$ behaviors with the smallest Hamming distances to the query signature.
\begin{equation}\label{E6}
     I_t = TopK (HammingDistance(h_i, h_{q_t})),
\end{equation}
The sub-sequence $S_{t}$ is obtained through $I_t$ when employing Simhash as the search algorithm:
\begin{equation}\label{E7}
    S_{t} = Search(S, q_{t};SimHash) = Gather(S, I_t).
\end{equation}

Let's compare the time complexity of computing the Hamming distance and the inner product of vectors. Assuming that multiplication is considered an atomic operation, the complexity of computing the inner product of two d-dimensional vectors is $O(d)$. 
In contrast, computing the Hamming distance between two m-dimensional signatures is approximated as $O(1)$ because this calculation can leverage the XOR operation. 
Therefore, our method significantly reduces the time complexity from $O(d)$ to $O(1)$. Additionally, when model training updates item embeddings, the locality-sensitive property of SimHash ensures that the new binary signatures align with the updated item embeddings. Therefore, our search method improves the computational efficiency and ensures accuracy.

\subsubsection{Local-Aware Search Unit (LASU)}
In this unit, we broaden our focus beyond a single item. We make the granularity of observation coarser by using the user's recent behavior sequence to construct the query. Based on this query, which contains recent behavior information, we obtain a subsequence that represents the user's short-term interest, referred to as local-aware interest. It is worth noting that we do not directly use the recent behavior sequence as the SBS, as it may contain noise such as accidental clicks. Additionally, the user's long-term behavior sequence may include other behaviors that align with the local-aware interest.

Specifically, we represent the user's recent behavior sequence as $S_r = [e_{L-J}, \dots, e_L]$, where $J$ is the length of recent behavior sequence. 
Using average pooling to construct queries is a good choice. Additionally, considering that 
$J$ is very small, we can further model $S_r$ to enhance the representation ability of the query with slightly increasing the computational cost.
Considering that each behavior is a carrier of current potential interest, we use a GRU to extract a series of hidden states 
$H_r \in \mathbb{R}^{J \times d}$ from $S_r$. We then combine $H_r$ and $S_r$ to get the query $q_l$. $Avg()$ is the average pooling.
\begin{equation}
    H_r = GRU(S_r),
\end{equation}
\begin{equation}
    q_l = Avg(H_r) + Avg(S_r),
\end{equation}

The query $q_l$ guides the retrieval of the relevant SBS $S_{local} \in \mathbb{R}^{K \times d}$ using the same SimHash method employed in the target-aware search unit as described in \crefrange{E4}{E7}: 
\begin{equation}
    S_{l} = Search(S, q_{l}; SimHash).
\end{equation}
Since $q_l$ mines the semantic information of the user's recent behaviors,
$S_l$ retrieved through $q_l$ align with the user's local-aware interest, which is a coarser-grained interest.

\subsubsection{Global-Aware Search unit (GASU)}
In this unit, we expand the focus to encompass the entire behavior sequence to consider the user's more global and stable interest. Constructing a query using the user's entire long-term behavior sequence is challenging. On the one hand, directly average pooling the long sequence may result in poor query representation due to its length and the noise within the sequence. On the other hand, constructing the query with more complex modeling of the long sequence cannot meet the strict time limits of an online system. Additionally, for new users with relatively short behavior sequences, 
it is difficult to obtain their global and stable preferences.


To address the above issues, we employ clustering methods to aggregate the item set $M$ into several clusters using pre-trained item embeddings. Then we map the target item to the nearest cluster centroid in order to construct a query, which effectively and approximately captures the global preferences. Additionally, leveraging the semantic information of pre-trained embeddings helps to partially solve the user cold start problem.

We adopt a simplified Density Peak Clustering~\cite{dpc} (DPC) algorithm that doesn't require multiple iterations or additional parameters, efficiently identifying clusters of arbitrary shapes based on density connectivity. DPC assumes that a cluster center is surrounded by low-density neighbors and maintains a relatively large distance from high-density points.
For each item, we calculate its density $\rho$ and the minimum distance $\delta$ to items with higher density. 
\begin{equation}
    \rho_i = exp(-\sum_{e_j \in M}||e_i - e_j||^2_2),
\end{equation}
\begin{equation}
    \delta_i = 
    \begin{cases}
        \text{min}_{j:\rho_j > \rho_i} \ ||e_i - e_j||_2 ,& \text{if} \ \exists j  \ \text{s.t.}\  \rho_j > \rho_i \\
        \text{max}_j \ ||e_i - e_j||_2 , & \text{otherwise}
    \end{cases},
\end{equation}
The cluster center score of an item is set to $\rho \times \delta$.
DPC selects $C$ items with the highest scores as cluster centers and assigns the remaining items to the closest cluster center.

In this unit, we construct the query $q_{g}$ using the aforementioned clustering algorithm and average pooling. The query $q_{g}$ guides the retrieval of the relevant SBS $S_{g} \in \mathbb{R}^{K \times d}$ using the same SimHash method as described in \crefrange{E4}{E7}: 
\begin{equation}
q_{g} = DPC(e_{t}) + Avg(S),
\end{equation}
\begin{equation}
    S_{g} = Search(S, q_{g}; SimHash).
\end{equation}
where $DPC()$ means assigning $e_{t}$ to the nearest cluster center.
Since the cluster assignment for each item is precomputed, the complexity of it is $O(1)$ during inference.
Since $q_g$ contains the semantic information of the user's global preferences, $S_g$ retrieved through $q_g$ align with the user's global-aware interest, which is the most coarse-grained interest.

\subsection{Behavior Sequence Refinement Module}
After obtaining the subsequences that contains the user's multi-granularity interests, we aim to further mine sequential and interactive information to more accurately learn the user's diverse interests. For convenience, we unify $S_t,S_l,S_g$ as $S_* \in \mathbb{R}^{K \times d}$.

\subsubsection{Target-Aware Position Encoding (TAPE)}
Previous work has rarely emphasized the order of behaviors in SBSs, possibly sorting them by relevance in the retrieval module. However, we believe that the sequential information within SBSs plays a crucial role, much like the original sequence, in helping us better understand and learn the user's interests. Therefore, we preserve the original sequence of behaviors in SBS and use position encoding to supplement the sequential relationship, compensate for the information loss of the retrieval module, and enhance the expressive power of $S_*$. We use the relative position of each behavior in $S_*$ to the target item to construct the position encoding.
Rather than the widely used chronological order encoding~\cite{SAS}, this method better emphasizes the attention on the target information. 
Specifically, we propose the Target-Aware Position Encoding $TAPE \in \mathbb{R}^{L \times d}$. We then convert the retrieved indices $I$ in equation~\eqref{E6} to target-aware indices $I'$. Finally, we obtain the target-aware position embeddings through a look-up process and incorporate them into $S_*$:
\begin{equation}
    S_* = S_* + lookup(TAPE, I')).
\end{equation}

\subsubsection{Multi-Head Fourier Transformer (MHFT)}
After applying the above TAPE, our model has obtained the sequential information in SBSs. 
To efficiently learn the interactive information with less additional computational cost, we leverage the efficient \textbf{Fourier Transformer}. 
According to the convolution theorem~\cite{C}, the product operator in the frequency domain is equivalent to global circular convolution in the time domain, which is able to extract interactive information and refine the sequence representation.
It also implies that complex convolution operations can be implemented through fast Fourier transform (FFT) with a time complexity of $O(KlogK)$ and multiplication operations~\cite{fnet, gnet}. 
Compared with the self-attention mechanism with a time complexity of $O(K^2)$, this method explores interactive information while improving efficiency.

Specifically, we perform Fourier transform along the sequence length dimension to transform $S_*$ to $X \in \mathbb{C}^{K \times d}$ which represents the frequency domain representation of $S_*$.
We then implement the convolution fusion of behaviors and mine interactive information by performing the element-wise multiplication with a complex parameter matrix $W \in \mathbb{C}^{K \times d}$~\cite{fmlp}.
Finally, we obtain the full convolution sequence representation $\widetilde{S_*} \in \mathbb{R}^{K \times d}$ through inverse fast Fourier transform (IFFT). The formulas are as follows:
\begin{equation}\label{E16}
    X = \mathcal{F} (S_*), \quad \widetilde{X} = W \odot X, \quad \widetilde{S_*} = \mathcal{F}^{-1} (\widetilde{X}),
\end{equation}

Additionally, inspired by the multi-head self attention mechanism, we further propose the \textbf{Multi-Head Fourier Transformer}.
To achieve better information fusion, we first use matrix multiplication instead of element-wise multiplication in equation~\eqref{E16}, using $W\in \mathbb{C}^{K \times d \times d}$ instead of $W\in \mathbb{C}^{K \times d}$. Then we shares weights $W\in \mathbb{C}^{ d \times d}$ among different tokens, to adaptively handle sequences of different lengths while reducing the parameter count.
To mine the interactive information from different subspaces, we divide $X \in \mathbb{C}^{K \times d}$ into $n$ parts $x_i \in \mathbb{C}^{K \times d/n}$ and diagonalize the weight $W\in \mathbb{C}^{ d \times d}$ into $n$ blocks $W_i \in \mathbb{C}^{d / n \times d / n}$. Each block of $X$ and the diagonal matrix is equivalent to a head of the multi-head self attention.
The diagonal weights can be computed in parallel.
We implement them through an MLP with an activation function $\sigma$ to introduce nonlinear relationships. The formulas are as follows:
\begin{equation}
    \widetilde{x}_i = \sigma(W_i x_i),
    \quad i=1,2,\dots, n,
\end{equation}
\begin{equation}
    \widetilde{X} = [~\widetilde{x}_1;~ \widetilde{x}_2;~ 
    \dots;~\widetilde{x}_n~],
\end{equation}

In this way, we further mine the interactive information from different subspaces adequately and efficiently.
Following previous work~\cite{SAS}, we also integrate skip connections~\cite{residual} and layer normalization~\cite{ln} to alleviate gradient vanishing and improve network stability. The formulas are as follows:
\begin{equation}
    \widetilde{X} = LayerNorm(\widetilde{X} + X),
\end{equation}

Through Target-Aware Position Encoding and the Multi-Head Fourier Transformer, we efficiently and comprehensively extract the sequential and interactive information in SBSs, resulting in refined sequence representations $\widetilde{S_*}$. Subsequently, we perform an average pooling aggregation operation to obtain the vector representation of the user's interest $E_* \in \mathbb{R}^{d}$:
\begin{equation}
    E_* = Avg(\widetilde{S_*}).
\end{equation}

\subsection{Multi-granularity Interest Activation Module}
After applying the above modules, we acquire the user's interest vectors $E_t, E_l, E_g$ at different granularities, capturing target-aware, local-aware, and global-aware interest, respectively. 
However, it is unclear how much each interest contributes to the click on the target item. 
Therefore, we utilize multi-head target attention to adaptively learn the influence of interests at different granularities on the target item, obtaining the user representation $E_u$:
\begin{equation}
    E_m = [E_t; E_l; E_g] \in \mathbb{R}^{3 \times d},
\end{equation}
\begin{equation}
    Q_i = W_i^Q e_t, \quad K_i = W_i^K E_m, \quad V_i = W_i^V E_m,
\end{equation}
\begin{equation}
    head_i = softmax(\frac{Q_i K_i^T}{\sqrt{d}})V_i,
\end{equation}
\begin{equation}
    E_u = Concat(head_1, \cdots, head_h)W^O.
\end{equation}
where $h$ donates the number of heads and $W_i^Q, W_i^K, W_i^V, W^O$ are all learnable parameters.

\subsection{Prediction Layer }
For the final CTR prediction, we simply concatenate the target item embedding $e_{t}$, user interest  representation $E_u$ and other feature vectors $E_{other}$ into the MLP layer:
\begin{equation}
    \hat{y} = sigmoid(MLP(e_{t}; E_u; E_{other})),
\end{equation}

As CTR prediction is typically formulated as a binary classification problem, we train our model under the Cross-Entropy loss:
\begin{equation}
L = - \frac{1}{N} \sum_{i=1}^{N} \left(y_i \cdot \log(p_i) + (1-y_i) \cdot \log(1-p_i)\right).
\end{equation}
where $N$ is the number of instances.

\subsection{Complexity Analysis}
We conduct a detailed analysis and comparison of the complexity of the two core modules of our MIRRN in this subsection. Assume that a user request with a batch size of $B$ and the multiplication is treated as an atomic operation.
\subsubsection{Multi-granularity Interest Retrieval Module (MIRM)}
The time complexity of query construction for TASU, LASU and GASU is approximately $O(1)$, $O(Jd^2)$ and $O(1)$ respectively. 
Given the slow training speed of GRU, we omit it in practical applications, allowing the time complexity for LASU to be regarded as $O(1)$. 
Then we employ the SimHash-based search method within these units, which requires the hash signatures of the user behaviors and the candidate items. Notably these hash signatures can be stored in a table before inference, enabling a lookup operation during inference to obtain the hash signatures of behaviors in $O(1)$ time. 
The hashing of candidate items has a time complexity of $O(Bmd)$ which  can be optimized to $O(Bmlogd)$~\cite{opt}.
Finally, the time complexity for filtering behaviors is $O(BL)$ as we can calculate the Hamming distance in $O(1)$ through the XOR operation.
Overall, the time complexity of MIRM is $O(BL + Bmlogd)$, a level that has been proven to be acceptable in real industrial scenarios~\cite{SDIM, ETA}. 

\subsubsection{Behavior Sequence Refinement Module (BSRM)} 
\label{sec:bsrm}
The time complexity of Target-Aware Position Encoding is $O(1)$ due to the lookup operation. And the time complexity of
Multi-Head Fourier Transformer (MHFT) is $O(Kd logK+ Kd^2/n)$. 
While MIRRN introduces additional time complexity due to BSRM, compared to other long-term user behavior modeling methods, the increase in inference time is relatively small since $K\ll L$. This will be further verified in Section~\ref{sec:time}.
Additionally, we analyze the complexity of MHFT in comparison to Fourier Transformer (FT) and Multi-Head Self Attention (MHSA), as shown in Table~\ref{tab:ca}.
Our findings reveal that FT and MHFT exhibit lower time complexity than the widely used MHSA. 
Moreover, MHFT has the advantage of requiring the fewest parameters among these three methods.

\begin{table}
  \caption{Time complexity and parameter count for MHSA, FT and MHFT.}
  \label{tab:ca}
  \begin{tabular}{l|l|l}
    \toprule
    Model & Time Complexity & Parameter count\\
    \midrule
    MHSA &  $K^2d + 4Kd^2$    &  $4d^2$\\
    FT  &  $Kd logK+ Kd$    &  $Kd$\\
    MHFT &  $Kd logK+ Kd^2/n$&   $d^2 / n$\\
    \bottomrule
\end{tabular}
\end{table}

\section{Experiments}
    
    
\subsection{Experimental Settings}
\subsubsection{Datasets}
We conduct comprehensive experiments on three real-word and large-scale datasets. The statistical details of these three datasets are presented in Table \ref{tab:dataset}.
\begin{itemize}[leftmargin=*,align=left]
    \item \textbf{Taobao\footnote{https://tianchi.aliyun.com/dataset/dataDetail?dataId=649}:
    } This dataset encompasses user behavior records from November 25 to December 3, 2017, drawn from one of China's largest e-commerce platforms. 
    \item \textbf{Alipay\footnote{https://tianchi.aliyun.com/dataset/dataDetail?dataId=53}:
    } The dataset is collected through Alipay, an online payment application. It covers users' online shopping behaviors spanning from July 1, 2015, to November 30, 2015.
    \item \textbf{Tmall\footnote{https://tianchi.aliyun.com/dataset/dataDetail?dataId=42}:
    } The dataset is provided by Alibaba Group which contains user behavior history on Tmall e-commerce platform from May 2015 to November 2015.
\end{itemize}

\subsubsection{Evaluation Metrics}
In our experiments, we use the AUC (Area Under ROC) score as the evaluation metric, which is the most commonly used metric in CTR tasks. AUC measures the ability of the model to prioritize positive samples over negative samples. A larger AUC value indicates better performance in predicting CTR.

\subsubsection{Baselines}
To evaluate the effectiveness of MIRRN, we conduct a comprehensive comparison with 14 state-of-the-art CTR prediction baseline models. We divide them into three categories. The first category is the feature interaction models, including DNN~\cite{youtube}, Wide\&Deep~\cite{wide}, and PNN~\cite{PNN}. The second category is the short-term behavior modeling models, including DIN~\cite{DIN}, DIEN~\cite{DIEN}, DSIN~\cite{DSIN}, DMIN~\cite{DMIN}, and CAN~\cite{CAN}. The third category is the long-term behavior modeling models, including MIMN~\cite{MIMN}, SIM Hard~\cite{SIM}, SIM Soft~\cite{SIM}, ETA~\cite{ETA}, SDIM~\cite{SDIM}, and TWIN~\cite{TWIN}.

\begin{table}
  \caption{The dataset statistics.}
  \label{tab:dataset}
  \begin{tabular}{crrrc}
    \toprule
    Dataset & Users & Items & Interaction & Avg. Seq. Length\\
    \midrule
    Taobao & 987,994 & 4,162,024 & 100,150,807 & 101\\
    
    Alipay & 498,308 & 2,200,291 & 35,179,371 & 70\\
    
    Tmall& 424,179 & 1,090,390 & 54,925,331 & 129 \\
    \bottomrule
\end{tabular}
\end{table} 

\subsubsection{Implementation Details}
For the three datasets, we adopt the same data preprocessing method as MIMN~\cite{MIMN}. All samples are divided into a training set (80\%), a validation set (10\%), and a test set (10\%) based on the click time of the target item.
For the short-term user behavior modeling model, we use the user's most recent 100 behaviors. For the long-term user behavior modeling model, we use the user's most recent 300 behaviors. For a fair comparison, except for the behavior sequence modeling module, our model and all baselines retain the same network structure.
It is worth noting that in our proposed MIRRN, there are three search units. To compare fairly with the two-stage retrieval models, we ensure that the total number of behaviors retrieved by MIRRN is the same as theirs.
The MLP structure of the final prediction layer of all models is 200×80×2, and the activation function is \textit{prelu} ~\cite{10.1109/ICCV.2015.123}. Each feature dimension is set to 16. The batch size for all models is 256. During model training, we use the Adam optimizer with a fixed learning rate of 0.001 and train for 1 epoch. To ensure the fairness of the experiment, we optimize all baselines according to their original papers and repeat the experiments 10 times to obtain a more stable evaluation.

\begin{table}
  \caption{Overall performance of various methods on three public datasets. "Imp." denotes the improvement rate of MIRRN over the strongest baseline. An asterisk (*) indicates statistical significance (with p<0.05) when comparing MIRRN to the best baseline results. }
  \label{tab:ex}
    \scalebox{1.1}{
  \begin{tabular}{l|c|c|c}
    \hline
    \multirow{2}{*}{Model} & \multicolumn{3}{c}{AUC}\\
    \cline{2-4}
     &       Taobao & Alipay & Tmall  \\
    \hline
    DNN(Avg-Pooling)& 0.8713& 0.8135& 0.8990\\ 
    Wide\&Deep&	0.8812&	0.8223&	0.9058\\ 
    PNN& 0.8989& 0.8347& 0.9236 \\
    \hline
    
    DIN&	0.9038&	0.8459&	0.9241\\
    DIEN&	0.9061&	0.8499&	0.9357\\
    DSIN&	0.9073& 0.8473&	0.9372\\
    DMIN&	0.9225&	0.8538& 0.9404\\
    CAN&	0.9247&	0.8500& 0.9344\\
    \hline

    MIMN&	0.9082&	0.8501&	0.9303\\
    SIM Hard&	0.9084&	0.8505&	0.9391\\
    SIM Soft&	0.9238&	\underline{0.8674}&	0.9455\\
    ETA&	0.9186&	0.8608&	0.9403\\
    SDIM&	0.9084&	0.8660& 0.9452\\
    TWIN&	\underline{0.9257}&	0.8656&	\underline{0.9459}\\
    \hline
    MIRRN&	\textbf{0.9319$^{*}$}&	\textbf{0.8791$^{*}$}&	\textbf{0.9507$^{*}$}\\
    Imp.& +0.67\% &+1.35\% &+0.51\%\\
    \hline
\end{tabular}
}
\end{table}

\subsection{Overall Performance}
In Table~\ref{tab:ex}, we present a thorough performance comparison on three public datasets. The following conclusions can be drawn:

(1). Our proposed MIRRN consistently outperforms all baselines across three public datasets, illustrating its effectiveness. Specifically, MIRRN achieves AUC improvements of
0.67\%, 1.35\%, and 0.51\% improvements over the strongest baseline on the Taobao, Alipay and Tmall datasets, respectively.

(2). The top two strongest baselines on the three datasets predominantly come from models on long-term behavior modeling. Among the three types of baselines, long-term behavior modeling models generally deliver the best performance, followed by short-term behavior modeling models. The experiments demonstrate that modeling user interest significantly enhances CTR model performance. Furthermore, leveraging more extensive user historical behaviors can lead to substantial performance gains.

(3). Among the short-term behavior modeling models, DMIN achieved the best performance on two datasets. DMIN emphasizes that users typically have multiple interests. It uses multi-head self-attention to implicitly extract multiple user interests.
The strong performance of DMIN highlights the effectiveness of learning users’ multiple interests, 
which is similar to the motivation behind our model.
Our proposed MIRRN further enhances performance by efficiently utilizing users' long-term behaviors and explicitly capturing diverse interests, thereby surpassing DMIN.

(4). Among the long-term behavior modeling baselines, SIM Soft and TWIN achieve strong results. SIM Soft uses the dot product of pre-trained embeddings as the relevance metric, while TWIN maintains consistency in relevance metrics during the two-stage interest modeling. However, both methods focus solely on the user's target-relevant behaviors during the first stage and lack the mining of relational information within subsequences during the second stage. In contrast, MIRRN considers the user's multi-granularity interests during the retrieval stage and effectively mines sequential and interactive information within subsequences, resulting in superior performance. These points will be further verified in the subsequent ablation study.

\begin{table}
  \caption{Ablation study of different search units.}
  \label{tab:ablation1}
  \begin{tabular}{l|ccc}
    \hline
    \multirow{2}{*}{Components} & \multicolumn{3}{c}{AUC} \\
     & Taobao & Alipay & Tmall \\
    \hline
    TASU         & 0.9139 & 0.8594 & 0.9400\\ 
    LASU         & 0.8910 & 0.8390 & 0.9238\\
    GASU         & 0.8664 & 0.8314 & 0.9041\\
    TASU \& LASU & 0.9284 & 0.8720& 0.9458\\
    TASU \& GASU & 0.9265 & 0.8680& 0.9462 \\
    LASU \& GASU & 0.8990 & 0.8489& 0.9268 \\
    \hline
    TASU \& LASU \& GASU & \textbf{0.9319} & \textbf{0.8791}& \textbf{0.9507}\\ 
  \hline
\end{tabular}
\end{table}
\begin{table}
  \caption{Ablation study of core components in MIRRN.}
  \label{tab:ablation2}
  \begin{tabular}{l|ccc}
    \hline
    \multirow{2}{*}{Method} & \multicolumn{3}{c}{AUC} \\
     & Taobao & Alipay & Tmall \\
    \hline
    MIRRN w/o BSRM & 0.9208 & 0.8686 & 0.9376\\
    MIRRN w/o MIAM & 0.9240 & 0.8710 & 0.9392\\
    \hline
    MIRRN w/o TAPE & 0.9278 & 0.8737 & 0.9436\\
    MIRRN w/o MHFT & 0.9256 & 0.8714 & 0.9394\\
    \hline
    MIRRN with MHSA  &0.9311 &\textbf{0.8803} &0.9488   \\
    MIRRN with FT    &0.9308 & 0.8784& 0.9476\\
    MIRRN with MHFT (Ours)  & \textbf{0.9319} & 0.8791& \textbf{0.9507}\\
    \hline

\end{tabular}
\end{table}

\subsection{Ablation Study}
We conduct comprehensive ablation experiments to verify the effectiveness of each module and support our ideas. 

(1). We assess the effectiveness of capturing multi-granularity interests by evaluating the three search units in MIRM, which are denoted as TASU, LASU and GASU. 
The experimental results, presented in Table~\ref{tab:ablation1}, are obtained with a constant total number of retrieved behaviors. Notably, 
the model using only TASU performs relatively well, which may explain why previous models focusing solely on target-aware interest still achieved notable results. In contrast, models using only LASU or GASU exhibit poor performance. Moreover, any model lacking TASU experiences a significant performance decline, underscoring the critical importance of TASU. Nevertheless, adding any of the search units consistently enhances performance, demonstrating the value of capturing users' diverse interests. Our proposed MIRRN, which employs all three search units, effectively captures interests at different granularities, leading to the best performance.

(2). We validate the effectiveness of BSRM and MIAM by separately removing each component. The results, as shown in Table~\ref{tab:ablation2}, reveal that removing either BSRM or MIAM results in a performance decrease. It indicates that both mining the relational information of behaviors and adaptively learning the impact of multi-granularity interests on the target item are crucial.

(3). We further verify the effectiveness of two key components, TAPE and MHFT, in BSRM,
which are responsible for mining sequential and interactive information respectively.
The results in Table~\ref{tab:ablation2} demonstrate that both sequential information in TAPE and interactive information in MHFT positively influence performance.

(4). We conduct comparative experiments on three methods for mining interactive information in sequences: FT, MHFT and MHSA. The results in Table~\ref{tab:ablation2} indicate that our proposed MHFT achieves performance comparable to MHSA and consistently outperforms FT.
MHFT even achieves the best results on two datasets. Notably, MHFT exhibits lower time complexity and requires fewer parameters as explained in Section \ref{sec:bsrm}. Therefore, MHFT demonstrates superiority among these three methods.




\subsection{Inference Time Analysis}
\label{sec:time}
\begin{figure}[t]
  \centering
  \includegraphics[width=\linewidth]{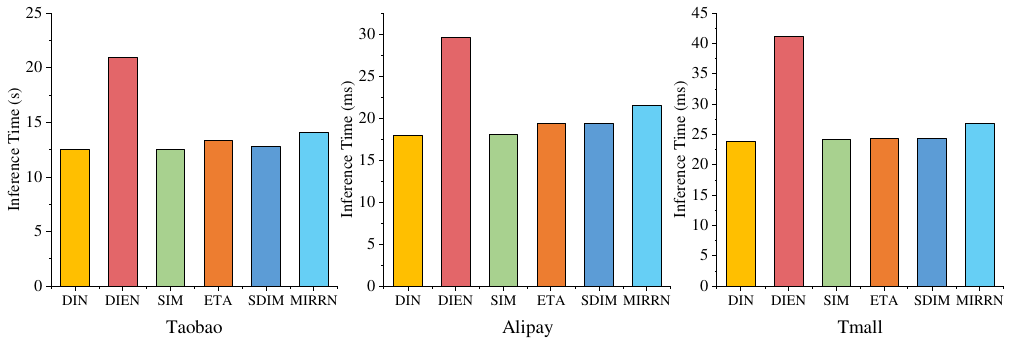}
  \caption{Inference time on the test set of three datasets.}
  \label{fig:time}
  \Description{}
\end{figure}
To demonstrate the practical value of our model, we compare the inference time across various models, including DIN, DIEN, SIM, ETA, SDIM, and MIRRN. 
As shown in Figure~\ref{fig:time}, SIM and SDIM exhibit relatively low inference times among the long sequence modeling methods. Our method introduces additional modules, leading to a slight increase in inference time. However, compared to the fastest baseline, our model only increases inference time by 12.0\%, 11.9\%, and 10.1\% across the three datasets, while significantly improving AUC by 1.45\%, 1.51\%, and 0.58\% over the strongest baseline.

\subsection{Online A/B Test}
\begin{figure}[t]
  \centering
  \includegraphics[width=\linewidth]{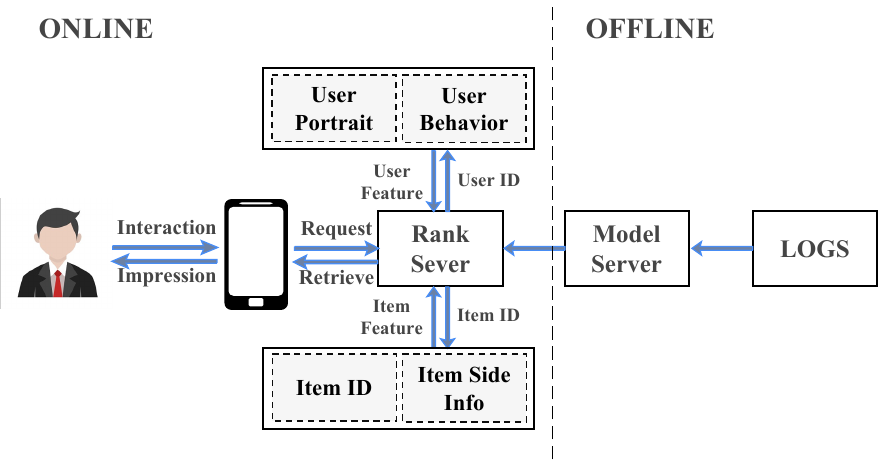}
  \caption{A simple illustration of the deployed playlist recommendation system in Huawei.}
  \Description{}
  \label{fig:abtest}
\end{figure}

We further conduct an online A/B test on the core playlist recommendation channel of the Huawei Music App, which is utilized by millions of users daily. Figure~\ref{fig:abtest} gives a simple illustration of the deployed system, which encompasses both the offline training and online inference processes. 
During the offline phase, the system logs hundreds of millions of user interactions to refine the prediction model, subsequently uploading the updated model to the online server for real-time user inference. 
In the online phase, upon receiving a new request, the online servers retrieve the relevant features and forward them to the ranking server, which then generates and presents a personalized recommendation list to users.

The baseline employed is a highly optimized deep CTR model. To ensure a fair comparison, identical input features are utilized across both the baseline model and our proposed model. An online A/B test is conducted over one week, involving 20\% of users selected at random. 
Of these users, half are designated to the experimental group while the remaining half are reserved as the control group. The evaluation is based on two metrics: the average number of listening songs (ANoLS) and the average time of listening songs (AToLS). The outcomes indicated that our model achieved an improvement of 1.32\% in ANoLS and 0.55\% in AToLS. Considering that an increment of 1\% in listening time is regarded as a substantial enhancement, our model realizes a significant business gain.




\section{Conclusion}

In this paper, we proposed the Multi-granularity Interest Retrieval and Refinement Network (MIRRN) for long-term user behavior modeling in CTR prediction. To the best of our knowledge, MIRRN is the first to successfully capture, refine, and integrate the user’s multi-granularity interests.
Specifically, we constructed multi-granularity queries, 
extracting subsequences that capture diverse interests. We then utilized Target-Aware Position Encoding and Multi-Head Fourier Transformer to efficiently and effectively mine sequential information and interactive information within these subsequences, obtaining the refined interest vectors.
Finally, we employed multi-head target attention to
adaptively learn the influence of different granular interests on the target item.
Experiments on three public datasets, along with an online A/B test validated the effectiveness of MIRRN. 
In the future, we plan to consider user interests at additional granularities tailored to different scenarios and explore more efficient retrieval methods to further reduce time complexity.

\section{Acknowledgments}
This work is supported by the National Natural Science Foundation of China (No. U23A20319, 62472394, and 62202443). And we thank MindSpore for the partial support of this work, which is a new deep learning computing framework. We would like to express our gratitude to FuxiCTR~\cite{fuxictr1, fuxictr2} for providing an open-source library for CTR prediction.

\bibliographystyle{ACM-Reference-Format}
\balance
\bibliography{content/references}

\end{document}